\documentclass[pop-hack,fleqn]{w-art-hack}
\usepackage{times}
\usepackage{w-thm}
\usepackage[]{graphicx}
\graphicspath{{img/}}
\usepackage{amsmath,amssymb}
\usepackage{cite}

\begin{document}
\DOIsuffix{theDOIsuffix}
\Volume{51}
\Issue{1}
\Month{01}
\Year{2003}
\pagespan{1}{}
\Reviseddate{}
\Accepteddate{}
\Dateposted{}


\title[Improving the purity of one- and two-qubit gates by AC fields]%
{Improving the purity of one- and two-qubit gates by AC fields}

\author[S. Kohler]{Sigmund Kohler\footnote{Corresponding author\quad
            E-mail:~{sigmund.kohler@physik.uni-augsburg.de}}}
\author[P. H\"anggi]{Peter H\"anggi\footnote{E-mail:
	peter.hanggi@physik.uni-augsburg.de}}
\address{Institut f\"ur Physik, Universit\"at Augsburg,
        Universit\"atsstra\ss e~1, D-86135 Augsburg, Germany}

\begin{abstract}
We investigate the influence of AC driving fields on the coherence
properties of one- and two-qubit gate operations.  In both cases, we
find that for suitable driving parameters, the gate purity improves
significantly.  A mapping of the time-dependent system-bath model to
an effective static model provides analytical results.  The resulting
purity loss compares favorably with numerical results.
\end{abstract}

\maketitle


\section{Introduction}

The experimental realization of one-qubit gates in solid state setups
\cite{Nakamura1999a, Vion2002a, Chiorescu2003a} and two-qubit gates in
ion traps \cite{Leibfried2003a, Schmidt-Kaler2003a} and Josephson
junctions \cite{Pashkin2003a} has demonstrated that these systems
provide remarkable coherence properties although the goal of $10^{-5}$
errors per gate operation \cite{Steane1998a} has not yet been accomplished
experimentally.  The unavoidable coupling to external degrees of
freedom and the thereby caused decoherence still presents a main
obstacle for the realization of a quantum computer.
Several proposals to overcome the ensuing decoherence have been put
forward, such as the use of decoherence free subspaces
\cite{Palma1996a, Duan1997a, Zanardi1997a, Lidar1998a, Beige2000a},
coherence-preserving qubits \cite{Bacon2001a}, quantum Zeno subspaces
\cite{Facchi2002a}, optimized pulse sequences \cite{Jirari2005a,
Spoerl2005a}, dynamical decoupling \cite{Viola1998a, Viola1999a,
Vitali2002a, Gutmann2005a, Falci2004a}, and coherent destruction of
tunneling \cite{Thorwart2000a, FonsecaRomero2004a}.  Theoretical studies of
decoherence of two-level systems have been
extended to gate operations in the presence of an environment in
Refs.~\cite{Loss1998a, Grifoni1999a, Governale2001a, Thorwart2002a,
Storcz2003a, Storcz2005a}.

A variety of suggestions towards coherence stabilization relies on the
influence of external fields.  One of the most prominent examples is
the application of a sequence of $\pi$-pulses that flip the sign of
the qubit-bath coupling operator resulting in a so-called dynamical
decoupling (DD) of the qubit from the bath \cite{Viola1998a,
Viola1999a, Vitali2002a, Gutmann2005a, Falci2004a,
FonsecaRomero2004a}.  A drawback of this scheme is the fact that it
eliminates only noise sources with a frequency below the repetition
rate of the pulses.  This clearly causes practical limitations.
However, these limitations may be circumvented by using a related
scheme based on continuous-wave driving, i.e.\ one with a harmonic
time-dependence, which allows higher driving frequencies.

A different proposal for coherence stabilization is to employ the physics
of the so-called coherent destruction of tunneling (CDT).  CDT has
originally been discovered in the context of tunneling in a driven
bistable potential \cite{Grossmann1991a, Grossmann1991b,
Grossmann1992a, Kayanuma1994a,
Grifoni1998a}.  There, it has been found that a particle which is
initially in the, say, left well of a symmetric bistable potential,
can be prevented from tunneling by the purely coherent influence of an
oscillating driving field.  This effect is stable against dissipation
in the sense that the AC field also decelerates the dissipative
transitions from the left to the right well \cite{Dittrich1993a,
Grifoni1995a, Hartmann2000a}.

A less frequently studied problem is the extension of coherence stabilization
protocols to systems that consist of two or more interacting qubits.
Such an interaction is essential for two-qubit operations which
represent an indispensable part of all quantum algorithms
\cite{Nielsen2000a, Makhlin2001a, Galindo2002a}.  For a CNOT gate based on an
isotropic Heisenberg interaction \cite{Loss1998a, Kane1998a}, it has
been proposed to stabilize coherence by applying a control field to
one of the qubits and thereby obtain an effective Ising interaction
which is less sensitive to the influence of environmental degrees of
freedom \cite{FonsecaRomero2005a}.  This scheme possesses the
beneficial properties that (i) it involves only intermediately large
driving frequencies that can lie well below the bath cutoff and (ii)
it does not increase the gate operation time.  Moreover, since the
driving field couples to the same coordinate as the quantum noise,
this coherence stabilization is distinctly different from the recently
measured dynamical decoupling of a spin pair from surrounding spin
pairs~\cite{Ramanathan2005a}.

In this work, we extend our previous studies on coherence
stabilization of one- \cite{FonsecaRomero2004a} and two-qubit
operations \cite{FonsecaRomero2005a} and, moreover, detail some
technical aspects.  In Section~\ref{sec:model}, we introduce a model
for two qubits coupled to a heat bath and derive a Bloch-Redfield
master equation to describe quantum dissipation and decoherence.  This
formalism is applied in Sections \ref{sec:one} and \ref{sec:two},
respectively, to the single qubit dynamics and to a
two-qubit gate operation.  In each case, we derive within a
rotating-wave approximation (RWA) analytical results for the decay of
the gate purity.  Thereby, the proper treatment of the qubit-bath
coupling is of crucial importance.  The computation of averages for
the ensemble of all pure states is deferred to the appendix.

\section{Quantum gate with bit-flip noise}
\label{sec:model}
We consider a pair of qubits described by the Hamiltonian \cite{Loss1998a,
Kane1998a, Nielsen2000a, Makhlin2001a, Galindo2002a}
\begin{equation}
  \label{H0}
  H_\mathrm{qubits}
 =\frac{1}{2} \sum_{j=1,2}
  \left( \Delta_j \sigma^z_j + \epsilon_j \sigma^x_j \right)
  + J\, \vec{\sigma}_{1}\cdot\vec{\sigma}_{2},
\end{equation}
with a qubit-qubit coupling of the Heisenberg type, where $j$ labels
the qubits.  In order to construct a quantum gate, the tunnel
splittings $\Delta_j$, the biases $\epsilon_j$, and the qubit-qubit
coupling $J$ have to be controllable in the sense that they can be
turned off and that their signs can be changed.

The bit-flip noise is specified by the system-bath Hamiltonian
\cite{Hanggi1990a, Leggett1987a}
\begin{equation}
\label{Htotal}
H=H_\mathrm{qubits}+H_\mathrm{coupl}+H_\mathrm{bath}
\end{equation}
where
\begin{equation}
\label{Hcoupl}
H_\mathrm{coupl} = \frac{1}{2} \sum_{j=1,2}\sigma_j^x \sum_\nu \hbar c_\nu
(a_{j\nu}^\dagger+a_{j\nu})
\end{equation}
denotes the coupling of qubit $j$ to a bath of harmonic oscillators with
frequencies $\omega_\nu$ described by the Hamiltonian
$H_\mathrm{bath} = \sum_{j\nu} \hbar\omega_\nu a_{j\nu}^\dagger
a_{j\nu}$ and the spectral density $I(\omega) =
\pi\sum_\nu c_\nu^2 \delta(\omega-\omega_\nu)$.
This coupling of each qubit to an individual bath represents a proper
model for sufficiently distant qubits \cite{Storcz2005a}.  Within the
present work, we consider the so-called ohmic spectral density
\begin{equation}
I(\omega) = 2\pi\alpha\omega \mathrm{e}^{-\omega/\omega_c}
\end{equation}
with the dimensionless coupling strength $\alpha$ and the cutoff frequency
$\omega_c$.
In order to complete the model, we specify the initial condition of
the Feynman-Vernon type, i.e., initially, the bath is in thermal
equilibrium and uncorrelated with the system, $\rho_\mathrm{tot}(t_0)
= \rho(t_0)\otimes R_\mathrm{bath,eq}$, where $\rho$ denotes the
reduced density operator of the two qubits and
$R_\mathrm{bath,eq}\propto\exp(-H_\mathrm{bath}/k_\mathrm{B}T)$ is the
canonical ensemble of the bath.

\subsection{Bloch-Redfield master equation}

If $\alpha k_\mathrm{B}T$ is smaller than the typical system energy and if the
dissipation strength is sufficiently small, $\alpha\ll 1$, the dissipative
system dynamics is well described within a Bloch-Redfield approach,
which is also referred to as Born-Markov approach.  There, one starts from the
Liouville-von Neumann equation $\mathrm{i} \hbar \dot \rho_\mathrm{tot} =
[H,\rho_\mathrm{tot}]$ for the total density operator and obtains by
standard techniques the master equation \cite{Kohler1997a}
\begin{align}
\dot\rho
= {} & -\frac{\mathrm{i}}{\hbar}[H_\mathrm{qubits},\rho]
  -\sum_j [\sigma_j^x, [Q_j(t),\rho ] ]
  - \sum_j [\sigma_j^x, \{ P_j(t),\rho \} ]
  \label{Born-Markov}
  \\
  \equiv {}& -\frac{\mathrm{i}}{\hbar}[H_\mathrm{qubits},\rho]
  - \Lambda(t)\rho
\label{Born-Markov2}
\end{align}
with the anti-commutator $\{A,B\}=AB+BA$ and
\begin{align}
  \label{Q(t)}
  Q_j(t) ={}& \frac{1}{4\pi}\int_0^\infty \mathrm{d}\tau \int_0^\infty
  \mathrm{d}\omega\,
  \mathcal{S}(\omega) \cos(\omega\tau) \widetilde\sigma_j^x(t-\tau,t) ,
  \\
  \label{P(t)}
  P_j(t) ={}& \frac{1}{4\pi}\int_0^\infty \mathrm{d}\tau \int_0^\infty
  \mathrm{d}\omega\,
  I(\omega) \sin(\omega\tau) \widetilde\sigma_j^x(t-\tau,t) .
\end{align}
Thus, the influence of the bath is determined by the Heisenberg
operators of the system, the spectral density $I(\omega)$ of the heat
baths, and the Fourier transformed
\begin{equation}
\mathcal{S}(\omega) = I(\omega)\coth(\hbar\omega/2k_\mathrm{B}T)
\end{equation}
of the symmetrically-ordered equilibrium auto-correlation function
$\frac{1}{2}\langle\{ \xi_j(\tau),\xi_j(0) \}\rangle_\mathrm{eq}$ of
the collective bath coordinate $\xi_j=\sum_\nu c_\nu
(a_{j\nu}^\dagger+a_{j\nu})$.  The notation $\widetilde X(t,t')$ is a
shorthand for the Heisenberg operator $U^\dagger(t,t') X U(t,t')$ with
$U$ being the propagator of the coherent system dynamics.
Note that $\mathcal{S}(\omega)$ and $I(\omega)$ are independent of $j$ due
to the assumption of two identical environments.
We emphasize that the particular form \eqref{Born-Markov} of the master
equation is valid also for an explicitly time-dependent qubit Hamiltonian.

\subsection{Purity decay}

The heat baths, whose influence is described by the second and third
term of the master equation \eqref{Born-Markov}, lead to decoherence,
i.e., the evolution from a pure state to an incoherent mixture.  This
process can be measured by the decay of the purity
$\operatorname{tr}(\rho^2)$ from the ideal value 1.  The gate purity
(frequently also referred to as ``purity'') $ \mathcal{P}(t) =
\overline{\operatorname{tr} (\rho^2(t))} $, which characterizes the
gate independently of the specific input, results from the ensemble
average over all pure initial states \cite{Poyatos1997a}.  For weak
dissipation, the purity is determined by its decay rate at initial
time,
\begin{equation}
  \label{puritydecay}
  \Gamma \equiv -\frac{\mathrm{d}}{\mathrm{d}t}\ln\mathcal{P}(t)\Big|_{t=0}
  = 2\,\overline{\operatorname{tr}(\rho\Lambda\rho)}
  = \frac{4}{N(N+1)} \sum_j \operatorname{tr}(\sigma_j^x Q_j) .
\end{equation}
In order to obtain the last expression, we have used the cyclic
property of the trace and performed the ensemble average over all pure
states as described in Appendix \ref{app:average}.  Here, $N$ is the
dimension of the system Hilbert space.
The purity loss rate $\Gamma$ represents a figure of merit for the
coherence of the quantum gate---ideally, it vanishes.
Interestingly enough, only the second term on the right-hand side of the
master equation \eqref{Born-Markov} contributes to the purity
decay.  This relates to the interpretation that the first term of the
master equation is
responsible for the coherent dynamics, while the second and third term
correspond to decoherence and relaxation, respectively.

The average over all pure states, which underlies the rate
\eqref{puritydecay}, may differ from the average over a discrete set
of initial states by a factor of the order unity, due to the particular
choice of a discrete ensemble: There, one commonly chooses ensembles
which are unsymmetric on the Bloch sphere \cite{Thorwart2002a,
Storcz2003a} or which do not include entangled states
\cite{Thorwart2002a, Storcz2003a, Storcz2005a}.

\subsection{Numerical solution}
\label{sec:numMethod}
The purity decay rate \eqref{puritydecay} by construction accounts only for
the behavior at initial time $t=0$.  Thus, for a more complete
picture, it is desirable to have the exact numerical solution of the
master equation \eqref{Born-Markov} at hand.  For studying the
influence of an external AC field, such a solution must
properly capture the case of a $T$-periodic system Hamiltonian.
An efficient scheme for
that purpose is a modified Bloch-Redfield formalism whose cornerstone is a
decomposition into the Floquet basis of the driven system
\cite{Kohler1997a}:  According to the Floquet theorem, the Schr\"odinger
equation of a driven quantum system with a Hamiltonian of the form
$H(t) = H(t+T)$ possesses a complete set
of solutions of the form $|\psi_\alpha(t)\rangle =
\exp(-\mathrm{i}\epsilon_\alpha t/\hbar) |\phi_\alpha(t)\rangle$.  The
so-called Floquet states $|\phi_\alpha(t)\rangle$ obey the time-periodicity
of the Hamiltonian and $\epsilon_\alpha$ denotes the so-called quasienergy.
The Floquet states are elements of an Hilbert space extended by a
$T$-periodic time coordinate and are computed from the eigenvalue
equation
\begin{equation}
\Big(H(t)-\mathrm{i}\hbar\frac{\mathrm{d}}{\mathrm{d}t}\Big)|\phi_\alpha(t)\rangle =
\epsilon_\alpha|\phi_\alpha(t)\rangle.
\end{equation}
In the Floquet basis
$\{|\phi_\alpha(t)\rangle\}$, the master equation
\eqref{Born-Markov} assumes the form
\begin{equation}
\dot\rho_{\alpha\beta} =
-\frac{\mathrm{i}}{\hbar}(\epsilon_\alpha-\epsilon_\beta) \rho_{\alpha\beta}
-\sum_{\alpha'\beta'} \Lambda_{\alpha\beta,\alpha'\beta'}(t)\,
\rho_{\alpha'\beta'}.
\end{equation}
Besides computational advantages, using the Floquet basis has the
benefit that it implies the numerically exact treatment of the
coherent system dynamics.  Thereby, one avoids artefacts like the
violation of equilibrium properties in the undriven limit
\cite{Petrov1996a, Petrov1996b, Novotny2002a, May2004a, Kohler2005a}.

For weak dissipation, we can replace within a
rotating-wave approximation $\Lambda(t)$ by its time average
\cite{Kohler1997a}.  Finally, we integrate the master equation to obtain
the dissipative propagator $\mathcal{W}_{\alpha\beta,\alpha'\beta'}$ which
provides the final state $\rho_{\mathrm{out},\alpha\beta} =
\sum_{\alpha'\beta'} \mathcal{W}_{\alpha\beta,\alpha'\beta'}
\rho_{\mathrm{in},\alpha'\beta'}$ and, thus,
allows one to evaluate all quantities of interest.
In particular, it is possible to compute for a pure initial state
$|\psi\rangle = \sum_\alpha c_\alpha|\phi_\alpha\rangle$, i.e.\ for
$\rho_{\mathrm{in},\alpha\beta} = c_\alpha c_\beta^*$, the final state
$\rho_\mathrm{out}$ which possesses the purity
\begin{equation}
\operatorname{tr}\rho_\mathrm{out}^2
= \sum_{\alpha\beta} \rho_{\mathrm{out},\alpha\beta}\,
                     \rho_{\mathrm{out},\beta\alpha}
= \sum_{\alpha\beta\alpha'\beta'\alpha''\beta''}
  \mathcal{W}_{\alpha\beta,\alpha'\beta'}\,
  \mathcal{W}_{\beta\alpha,\beta''\alpha''}\,
  c_{\alpha'} c_{\beta'}^* c_{\beta''} c_{\alpha''}^* \,.
\end{equation}
In order to average over all pure initial states, we employ Eq.\
\eqref{c4} derived in the Appendix to obtain the gate purity
\begin{equation}
\mathcal{P}_\mathrm{out} = \frac{1}{N(N+1)}\sum_{\alpha\beta\alpha'\beta'} \Big(
\mathcal{W}_{\alpha\beta,\alpha'\beta'}\mathcal{W}_{\beta\alpha,\beta'\alpha'} +
\mathcal{W}_{\alpha\beta,\alpha'\alpha'}\mathcal{W}_{\beta\alpha,\beta'\beta'} \Big) .
\end{equation}
We emphasize that this result is independent of the particular choice of
the basis.

\section{Coherence stabilization for single qubits}
\label{sec:one}

\begin{vchfigure}[t]
  \includegraphics[scale=.5]{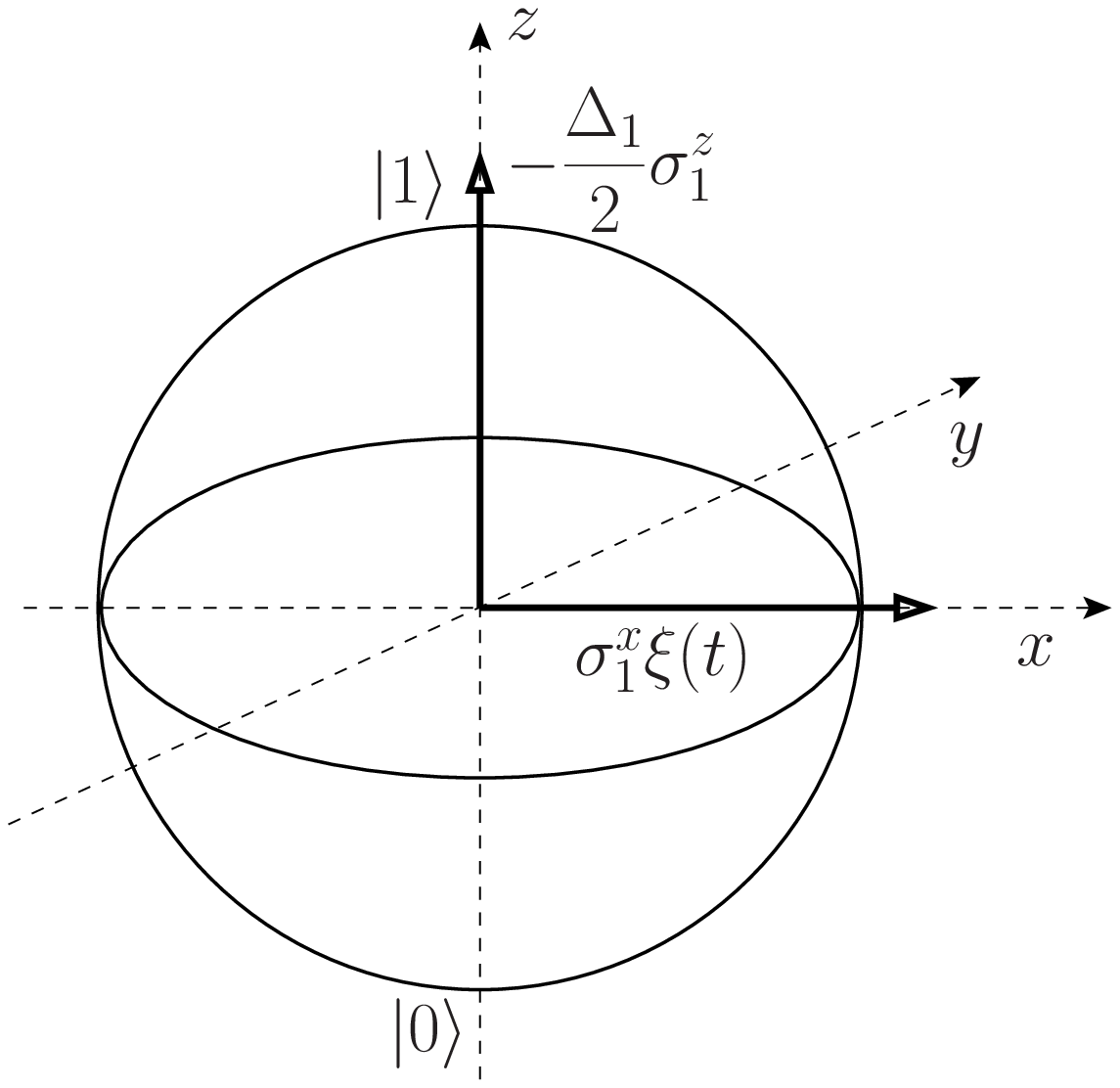}%
  \vchcaption{Bloch sphere representation of the single-qubit
  Hamiltonian \eqref{Hqubit1} and the bath coupling studied in
  Sec.~\ref{sec:one}.  \label{fig:bloch}}
\end{vchfigure}%
During the stage of single qubit operations, i.e., for $J=0$, both
qubits together with the respective bath evolve independently of each
other.  Thus,
it is sufficient to focus on qubit 1 with the Hilbert space dimension
$N=2$.  We restrict ourselves to an operation with $\epsilon_1=0$ and
the fixed tunnel splitting $\Delta_1 > 0$, i.e.\ to the Hamiltonian
\begin{equation}
\label{Hqubit1}
H_\mathrm{qubit1} = \frac{\Delta_1}{2} \sigma_1^z .
\end{equation}
This situation is
characterized by the fact that the bath couples to a system operator
which is different from the qubit Hamiltonian.  In the Bloch sphere
representation sketched in Fig.~\ref{fig:bloch}, the qubit Hamiltonian
\eqref{Hqubit1} and the bath coupling correspond to orthogonal vectors.
We start out by bringing the master equation \eqref{Born-Markov} into
a more explicit form by inserting the Heisenberg operator
\begin{equation}
\tilde\sigma_1^x(t-\tau,t)
= \sigma_1^x\cos\frac{\Delta_1\tau}{\hbar}
 +\sigma_1^y\sin\frac{\Delta_1\tau}{\hbar},
\end{equation}
which is readily derived from its definition together with the qubit
Hamiltonian \eqref{H0}.  Performing the integration over $\tau$ and
$\omega$ and neglecting renormalization effects, which are small provided
that $\alpha\ln (\omega_c/\Delta_1)\ll 1$, yields for $\Delta_1\ll\omega_c$,
the Markovian master equation
\begin{equation}
\label{ME1}
\dot\rho = -\frac{\mathrm{i}}{\hbar}[H_\mathrm{qubit1},\rho]
-\frac{\mathcal{S}(\Delta_1/\hbar)}{8}[\sigma_1^x,[\sigma_1^x,\rho]]
+\mathrm{i}\frac{\pi\alpha\Delta_1}{4}[\sigma_1^x,\{\sigma_1^y,\rho\}] .
\end{equation}
Then the purity decay \eqref{puritydecay} is readily evaluated to read
\begin{equation}
\Gamma_0 = \frac{1}{6}\mathcal{S}(\Delta/\hbar) ,
\end{equation}
where the subscript ``$0$'' refers to the absence of any AC field.

A possible coupling to an external driving field might have any
``direction'' $\vec n$ on the Bloch sphere (cf.\
Fig.~\ref{fig:bloch}), i.e.\ be proportional to $\vec
n\cdot\vec\sigma_1$.
Herein, we restrict ourselves to the cases parallel to the bath
coupling and parallel to the static Hamiltonian, i.e., to a field that
couples to $\sigma_1^z$ and $\sigma_1^x$, respectively.

\subsection{Dynamical decoupling by harmonic driving}
\label{DD}

The first option is to act on the qubit with a driving of the form
\begin{equation}
\label{HFDD}
H_\parallel = \frac{A}{2}\sigma_1^z \cos(\Omega t)
\end{equation}
``parallel'' to the static Hamiltonian.  This relates to a
recently proposed mechanism for coherence stabilization, namely the so-called
dynamical decoupling (DD) \cite{Viola1998a, Viola1999a,
Vitali2002a, Gutmann2005a, Falci2004a}.  This scheme employs sequences of
$\pi$-pulses that flip the sign of the operator $\sigma_1^x$ which
couples the qubit to the bath.  The basic idea dates back to
the suppression of spin diffusion in nuclear magnetic resonance
experiments \cite{Carr1954a,Haeberlen1968a} and
by now is an established technique in that area \cite{Slichter1990a}.
In the present case where the bath couples to the operator
$\sigma_1^x$ [cf.\ Eq.~\eqref{Hcoupl}], such a transformation is
e.g.\ induced by the Hamiltonian $\hbar\omega_\mathrm{R}\sigma_1^z$
for a pulse duration $\pi/\omega_\mathrm{R}$.  Since the corresponding
propagator is a function of the qubit Hamiltonian, the coherent
dynamics is not altered.
Besides the prospective benefits of such a control scheme, there is also a
number of possible drawbacks that the application of $\pi$-pulses might
cause: For a driven system, there is always the possibility of unwanted
off-resonant transitions \cite{Tian2000a}, especially in the case of
ideal rectangular
pulses.  A more practical limitation is the fact that only noise with
frequencies below the pulse repetition rate can be eliminated in such a way.
These disadvantages can be overcome partially by applying a continuous wave
version of the dynamical decoupling scheme, i.e.\ a driving of the
form \eqref{HFDD} for which the available frequency range is larger.

For the computation of the coherence properties, we use the fact that
$H_\parallel(t)$ commutes with the static qubit Hamiltonian
\eqref{Hqubit1} and, consequently, the propagator for the driven qubit
can be computed exactly reading
\begin{equation}
U(t,t')
= \exp\left(-\mathrm{i}\frac{A}{2\hbar\Omega}[\sin(\Omega t)- \sin(\Omega t')]
  \sigma_1^z\right)
  \exp\left(-\frac{\mathrm{i}}{\hbar}\Delta_1\sigma_1^z(t-t')\right) .
\end{equation}
We have written the propagator in a form that is suitable for
simplifying the master equation \eqref{Born-Markov}.
Inserting this into the expression \eqref{Q(t)} results for
$\Delta_1\ll\Omega$ in the effective coupling operator
\begin{equation}
\label{QDD}
Q_\mathrm{DD} = \frac{1}{2}
\left(J_0^2(A/\hbar\Omega) \mathcal{S}(\Delta_1/\hbar)
+2 \sum_{n=1}^\infty J_n^2\left({A}/{\hbar\Omega}\right)
\mathcal{S}(n\Omega) \right)\sigma_1^x .
\end{equation}
In order to derive this expression, we have decomposed the exponentials of
the trigonometric functions into a Fourier series using the identity
$\exp[\mathrm{i}x\sin(\Omega t)]=\sum_k J_k(x)
\exp(\mathrm{i}k\Omega t)$, where $J_k$ is the $k$th order Bessel function
of the first kind \cite{Gradshteyn1994a}.  The effective coupling
operator $Q_\mathrm{DD}$ is proportional to $\sigma_1^x$ and, thus, the
master equation is again of the form \eqref{Born-Markov}.  The only
difference is that the dissipative terms have acquired the prefactor
\begin{equation}
\label{etaDD}
\eta_\mathrm{DD}
= J_0^2\left(A/\hbar\Omega\right)
 +2 \sum_{n=1}^\infty \frac{n\hbar\Omega}{\Delta_1}
 \frac{\tanh(\Delta_1/2k_\mathrm{B}T)}{\tanh(n\hbar\Omega/2k_\mathrm{B}T)}
 \mathrm{e}^{-n\Omega/\omega_c} J_n^2\left(A/{\hbar\Omega}\right)
\end{equation}
which modifies the purity decay rate accordingly,
$ \Gamma_\mathrm{DD} = \eta_\mathrm{DD}\Gamma_0$.
Equation \eqref{etaDD} allows for the interpretation that now the
decoherence rate depends on the spectral density of the bath at
multiples of the driving frequency $\Omega$ which may be larger than the cutoff
frequency $\omega_c$.  The $\pi$-pulses applied in the original
version \cite{Viola1998a} of dynamical decoupling, correspond for a
continuous driving to a field amplitude that is adjusted such that
$A/\hbar\Omega$ equals the first zero of the Bessel function $J_0$,
i.e.\ it assumes a
value $2.404825\ldots$.  Then only the sum in Eq.~\eqref{etaDD}
contributes to the decoherence rate $\Gamma_\mathrm{DD}$.  If now the
driving frequency is larger than the cutoff of the spectral density,
$\Omega>\omega_c$, decoherence is considerably reduced: For
low temperatures, $k_\mathrm{B}T\ll\Delta_1$, the hyperbolic tangent in the
factor \eqref{etaDD} becomes unity and each contribution is weighted
by a possibly large factor $n\hbar\Omega/\Delta_1$.  In the
high-temperature limit $k_\mathrm{B}T\gg\hbar\Omega$, we use $\tanh(x)\approx
x$ and find that the dependence of the prefactor on $n\Omega$ cancels.
This means that the dynamical decoupling scheme is especially useful
for high temperatures.  The physical reason for this is that the
driving shifts the qubit dynamics towards high frequencies where the
thermal occupation of the bath modes is negligible.

\begin{figure}[t]
\centering
\includegraphics{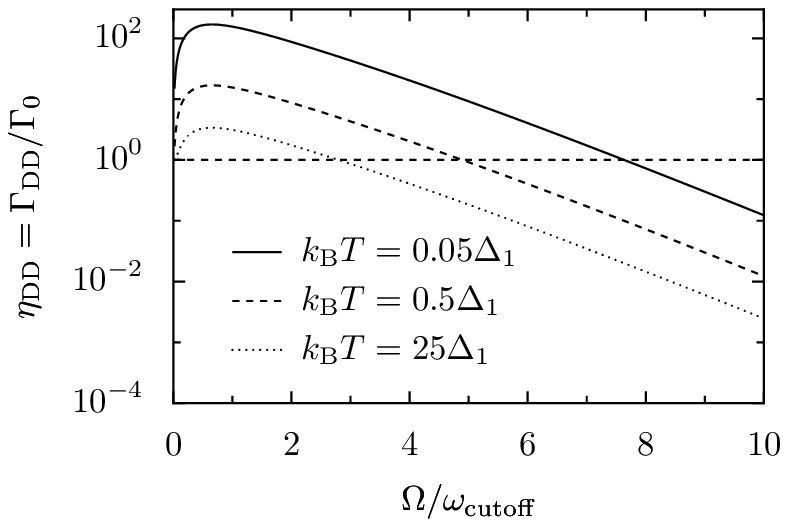}
\caption{Decoherence reduction by dynamical decoupling,
$\eta_\mathrm{DD}$, as a function of the driving
frequency for various temperatures.  The
cutoff frequency is $\omega_c=500\Delta_1/\hbar$, $A/\hbar\Omega=2.4$, and
the dissipation strength is $\alpha=0.01$.  The horizontal line marks
the value 1 below decoherence is lower than in the static case.
\label{fig:DD}}
\end{figure}

Figure \ref{fig:DD} compares the coherence stabilization
$\eta_\mathrm{DD}$ as a function of the driving frequency for
$A/\hbar\Omega=2.4$, i.e.\ close to a zero of the Bessel function
$J_0$.  It reveals that for driving frequencies well below the cutoff,
the driving rather spoils the coherence.  This improves with
increasing driving frequency and, finally, for a high-frequency
driving, $\eta_\mathrm{DD}$ becomes much smaller than unity
corresponding to a significant coherence stabilization.  The data
demonstrate the particular usefulness of dynamical decoupling at high
temperatures.

\subsection{Coherent destruction of tunneling}
\label{CDT}

Our second example under consideration is a driving field
\begin{equation}
\label{HF:CDT}
H_\perp(t) = f(t)\sigma_1^x ,
\end{equation}
where the energy $f(t)$ is a $2\pi/\Omega$-periodic function of time
with zero mean.  The field couples to the qubit by the same operator
$\sigma_1^x$ as the bath, i.e., ``perpendicular'' to the static
Hamiltonian \eqref{Hqubit1}.  Thus, it commutes with the qubit-bath
coupling but not with the static Hamiltonian.  Such a time-dependent
field causes already interesting effects for the coherent qubit
dynamics that we will briefly review before discussing decoherence.

For that purpose, we derive within a rotating-wave approximation (RWA)
analytical expressions for both the coherent propagator $U(t,t')$ and
the purity decay~\eqref{puritydecay}.  We start out by transforming
the total Hamiltonian into a rotating frame with respect to the
driving via the unitary transformation
\begin{equation}
\label{Uac}
U_\mathrm{AC}(t) = \mathrm{e}^{-\mathrm{i}\phi(t)\sigma_1^x},\quad
\phi(t) = \frac{1}{\hbar} \int_0^t \mathrm{d}t'\,f(t') .
\end{equation}
This yields the likewise $2\pi/\Omega$-periodic interaction-picture
Hamiltonian
\begin{align}
\widetilde H_\mathrm{qubit1}(t)
={}& U_\mathrm{AC}^\dagger(t) H_\mathrm{qubit1} U_\mathrm{AC}(t)
\\
={}& \frac{\Delta_1}{2}\left\{
   \sigma_1^z\cos\left(\frac{A}{\hbar\Omega}\sin(\Omega t)\right)
  +\sigma_1^y\sin\left(\frac{A}{\hbar\Omega}\sin(\Omega t)\right)
  \right\}
\label{Hinteraction}
\end{align}
and the S-matrix $S(t,t') = U_\mathrm{AC}^\dagger(t) U(t,t')
U_\mathrm{AC}(t')$.  The
corresponding Schr\"odinger equation cannot be integrated exactly
since $\widetilde H_\mathrm{qubit1}(t)$ does not commute with itself at
different times and, thus, time-ordering has to be taken into account.
We restrict ourselves to an approximate solution and neglect
corrections of the order $\Delta_1^2$.  Within this approximation, the
propagator is simply given by the exponential of the integral of the
time-dependent interaction-picture Hamiltonian.  This is equivalent
to replacing \eqref{Hinteraction} by its time-average
\begin{equation}
  \bar H_\mathrm{qubit1}
  \equiv \langle\widetilde H_\mathrm{qubit1}(t)\rangle_{2\pi/\Omega}
  = \frac{\Delta_\mathrm{eff}}{2}\sigma_1^z ,
\end{equation}
where $\langle\ldots\rangle_{2\pi/\Omega}$ denotes the time-average
over the driving period.  This RWA approximation to the driven qubit
Hamiltonian is of the same form as the original static
Hamiltonian~\eqref{H0}, but with the tunneling matrix element
being renormalized according to
\begin{equation}
\label{Delta.eff}
\Delta_1 \to \Delta_\mathrm{eff}
=\langle\cos[2\phi(t)]\rangle_{2\pi/\Omega} \Delta_1 ,
\end{equation}
Consequently, we find $S(t,t') =
\exp\{-\mathrm{i}\bar H_\mathrm{qubit1}(t-t')/\hbar\}$, such that
within RWA, the entire propagator for the qubit in the Schr\"odinger
picture reads
\begin{equation}
\label{U(t,t')}
U(t,t')
= \mathrm{e}^{-\mathrm{i}\phi(t)\sigma_1^x}\,
  \mathrm{e}^{-\mathrm{i} \bar H_\mathrm{qubit1}(t-t')/\hbar}\,
  \mathrm{e}^{\mathrm{i}\phi(t')\sigma_1^x} .
\end{equation}

Of particular interest are now driving parameters for which the
effective tunnel splitting \eqref{Delta.eff} and, thus,
$\bar H_\mathrm{qubit1}$ vanish.  Then, the one-period propagator $U(t+T,t)$
becomes the identity [recall that $U_\mathrm{AC}$ is
$2\pi/\Omega$-periodic and, thus, $U_\mathrm{AC}(2\pi/\Omega)=
U_\mathrm{AC}(0) = \mathbf{1}$].
This implies that the long-time dynamics is suppressed.
The dynamics within the driving period requires a closer look at the
$2\pi/\Omega$-periodic contribution $U_\mathrm{AC}(t)$: For an initial
preparation in an eigenstate of $\sigma_1^x$, it provides only a global
phase, such that the dynamics as a whole is suppressed also within the
driving period.  This effect of suppressing the time-evolution by the
purely coherent influence of an external field has been investigated
first in the context of driven tunneling \cite{Grossmann1991a,
Grossmann1991b} and is named ``coherent destruction of tunneling''
(CDT).  Therefore, we will refer to a driving of the form \eqref{HF:CDT} as
``CDT driving'' despite the fact that we also consider working points
at which the coherent dynamics is not completely suppressed.  Note
that for a preparation other than an eigenstate of $\sigma_1^x$, the
periodic propagator $U_\mathrm{AC}(t)$ will still cause a non-trivial
dynamics within the driving period.

Let us now turn to the influence of the CDT driving \eqref{HF:CDT} on
quantum dissipation and decoherence.  For that purpose, we have to
evaluate the operator $Q_j$ contained in the master equation
\eqref{Born-Markov}.  Inserting the RWA propagator \eqref{U(t,t')} into
\eqref{Q(t)}, we obtain after some algebra the result
\begin{equation}
\label{Q_CDT}
Q_\mathrm{CDT}
= \frac{1}{8}\mathcal{S}(\Delta_\mathrm{eff}/\hbar) \sigma_1^x .
\end{equation}
Thus, the master equation again is of the same form as in the undriven
case, Eq.~\eqref{Born-Markov}, but the generator of the dissipative
dynamics is modified by the factor
\begin{equation}
\label{etaCDT}
\eta_\mathrm{CDT}
= \frac{\mathcal{S}(\Delta_\mathrm{eff}/\hbar)}{\mathcal{S}(\Delta_1/\hbar)}
= \frac{\coth(\Delta_\mathrm{eff}/2k_\mathrm{B}T)}%
       {\coth(\Delta_1/2k_\mathrm{B}T)}
  \langle\cos[2\phi(t)]\rangle_{2\pi/\Omega}.
\end{equation}
Consequently, the purity decay becomes $\Gamma_\mathrm{CDT} =
\eta_\mathrm{CDT}\Gamma_0$.  Since the spectral density
$\mathcal{S}(\omega)$ increases monotonically with the frequency
$\omega$ and, moreover, the Bessel function $J_0(x)\leq 1$, the CDT
driving---in clear contrast to the dynamical decoupling---never enhances
dissipation and decoherence.

In the high-temperature limit $k_\mathrm{B}T\gg\Delta_\mathrm{eff}$, we employ
the approximation $\coth(x)\approx 1/x$ which implies that for an
ohmic bath, $\mathcal{S}(\Delta/\hbar) \approx 4\pi\alpha k_\mathrm{B}T/\hbar$
is independent of the tunnel splitting.  Consequently,
$\eta_\mathrm{CDT}\approx 1$, i.e., the purity decay is essentially
unchanged.

In the opposite limit of low temperatures, $k_\mathrm{B}T \ll
\Delta_\mathrm{eff}$, the argument of the hyperbolic cotangent is
large such that $\coth(x)\approx 1$.  Then $\eta_\mathrm{CDT} =
\Delta_\mathrm{eff}/\Delta_1 = \langle\cos[2\phi(t)]\rangle_{2\pi\Omega}
\leq 1$.  This reduction of decoherence is brought about by the fact
that the driving \eqref{HF:CDT} decelerates the long time dynamics of
the qubit.  Thereby, the frequencies which are relevant for the
decoherence are shifted to a range where the spectral density of the
bath is lower.  Consequently, the influence of the bath is diminished.

An important special case is that of a harmonically time-dependent driving
field, $f(t)= \frac{1}{2}A\cos(\Omega t)$.  Then, the time-average in
the effective matrix element \eqref{Delta.eff} can be expressed by a
zeroth-order Bessel function of the first kind such that
$\Delta_\mathrm{eff} = J_0(A/\hbar\Omega)\Delta_1$.  Consequently, we
find that at sufficiently low temperatures decoherence is reduced by a
factor $\eta_\mathrm{CDT} = J_0(A/\hbar\Omega)$.  These results imply
that for an ohmic bath, the coherent dynamics is slowed down by the
same factor as the decoherence, cf.\ Eqs.~\eqref{Delta.eff} and
\eqref{etaCDT} in the low-temperature limit.  Thus, if for a specific
application, the figure of merit is the number of coherent
oscillations, the present coherence stabilization scheme may
therefore not prove very useful.

\section{Coherence stabilization for a CNOT gate}
\label{sec:two}

Logic operations in both quantum and classical computers require that
the time evolution of a qubit (or a bit, respectively) depends on the
state of another qubit.  The physics behind such processes is a
non-linear interaction between two qubits which in many
implementations is of the Heisenberg type \cite{Loss1998a, Kane1998a}
\begin{equation}
\label{Hheisenberg}
H_\mathrm{Heisenberg} = J \vec\sigma_1\cdot\vec\sigma_2
\end{equation}
as assumed in our model Hamiltonian \eqref{H0}.  In the context of
decoherence, the question arises whether it is possible to stabilize
coherence by a proper AC field while maintaining the desired
non-linear time evolution.

\begin{vchfigure}
\includegraphics{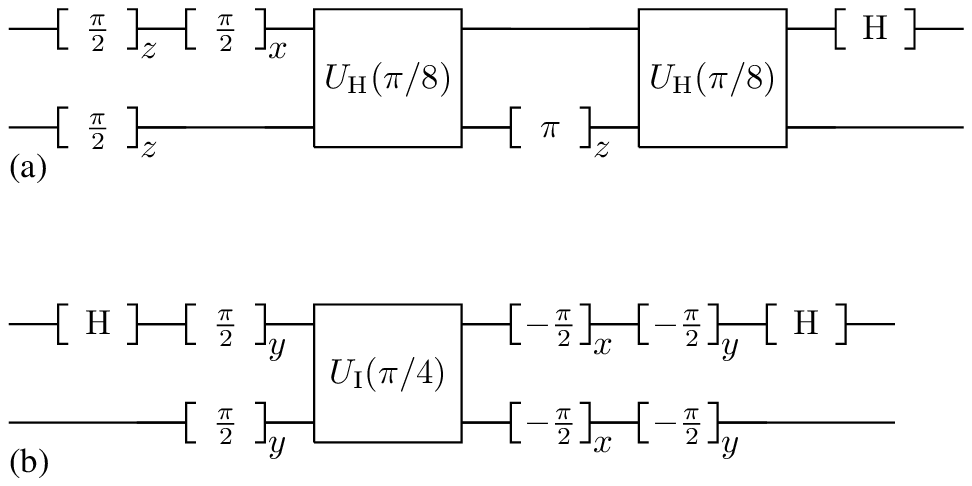}
\vchcaption{Realizations of a CNOT operation for different types of
qubit-qubit interaction \cite{Nielsen2000a, Makhlin2001a, Galindo2002a}.
(a) Heisenberg interaction providing the transformation \eqref{UH} and
(b) Ising coupling along the $x$-direction corresponding to
\eqref{UI}.
The symbol $[\mathrm{H}]$ denotes the Hadamard operation and
$[\phi]_n$ the rotation of the respective qubit around the axis $n$ by
an angle $\phi$.
\label{fig:cnot}}
\end{vchfigure}
The qubit Hamiltonian \eqref{H0} together with the bath coupling
\eqref{Htotal}, allows one to implement the CNOT operation sketched in
Figure \ref{fig:cnot}b \cite{Loss1998a, Nielsen2000a, Makhlin2001a,
Thorwart2002a, Galindo2002a, Schuch2003a}.  Apart from single qubit
operations, it consists of the propagator
\begin{equation}
\label{UH}
U_\mathrm{H}(\varphi) = \exp(-\mathrm{i} \varphi\,
\vec{\sigma}_1 \cdot\vec{\sigma}_2 )
\end{equation}
for the Heisenberg qubit-qubit interaction \eqref{Hheisenberg} which
in total acts for a time $t_J = \pi\hbar/4J$ such that $\varphi =
\pi/4$.
For single qubit operations, we have discussed in Sec.~\ref{sec:one}
that pulse sequences \cite{Viola1998a, Viola1999a, Vitali2002a, Gutmann2005a,
Falci2004a} and harmonic driving fields \cite{FonsecaRomero2004a} can
suppress decoherence.  Therefore, we focus here on decoherence during
the stage of the qubit-qubit interaction and, thus, take as a working
hypothesis that the coherence of one-qubit operations can be
stabilized ideally.  Then the remaining decoherence takes place during
the qubit-qubit interaction time $t_J$.

\subsection{Heisenberg vs.\ Ising coupling}
\label{sec:ising}

Before discussing the influence of a driving field, we provide for
later reference the results for the purity decay in the static situation.
Therefore, we need to evaluate $\operatorname{tr}(\sigma_j^x Q_j)$
where
\begin{equation}
Q_j = \frac{1}{4\pi} \int_0^\infty \mathrm{d}t \int_0^\infty
\mathrm{d}\omega\, \mathcal{S}(\omega) \cos(\omega\tau)\,
\mathrm{e}^{\mathrm{i}H_\mathrm{Heisenberg}\tau/\hbar} \sigma_j^x \,
\mathrm{e}^{-\mathrm{i}H_\mathrm{Heisenberg}\tau/\hbar} .
\end{equation}
This calculation is most conveniently performed in the basis of the
total (pseudo) spin $\vec L = \frac{1}{2}(\vec\sigma_1 +
\vec\sigma_2)$ because $H_\mathrm{Heisenberg} =J(2\vec L^2-3)$, where
$\vec L^2$ possesses the eigenvalues $\ell(\ell+1)$, $\ell=0,1$.  For
$\ell=0$, the corresponding eigenstate is the singlet state
$(|01\rangle-|10\rangle)/\sqrt{2}$ with energy $\epsilon_0=-3J$, while
for $\ell=1$, one finds the triplet $|00\rangle$,
$(|01\rangle+|10\rangle)/\sqrt{2}$, $|11\rangle$ with $\epsilon_1=J$.
After evaluating the matrix elements of $\sigma_j^x$ and $Q_j$, we
finally arrive at the purity decay rate
\begin{equation}
\label{GammaH}
\Gamma_\mathrm{Heisenberg}
  = \frac{2}{5}\{ \mathcal{S}(0) + \mathcal{S}(4J/\hbar) \}
\end{equation}
which is sketched in Fig.~\ref{fig:ising}.  In the derivation,
we have ignored Lamb shifts and defined
\begin{equation}
\mathcal{S}(0) \equiv
\lim_{\omega\to 0} \mathcal{S}(\omega) = \frac{4\pi}{\hbar}\alpha
k_\mathrm{B}T.
\end{equation}
In particular, we find that for low temperatures, $k_\mathrm{B}T \lesssim J$,
decoherence is dominated by the term $\mathcal{S}(4J/\hbar)$ such that
$\Gamma \approx 16\pi\alpha J/5\hbar$.  This part reflects the
influence of the so-called quantum noise which is
temperature-independent and, thus, cannot be reduced by cooling the
environment.
\begin{vchfigure}[t]
\includegraphics{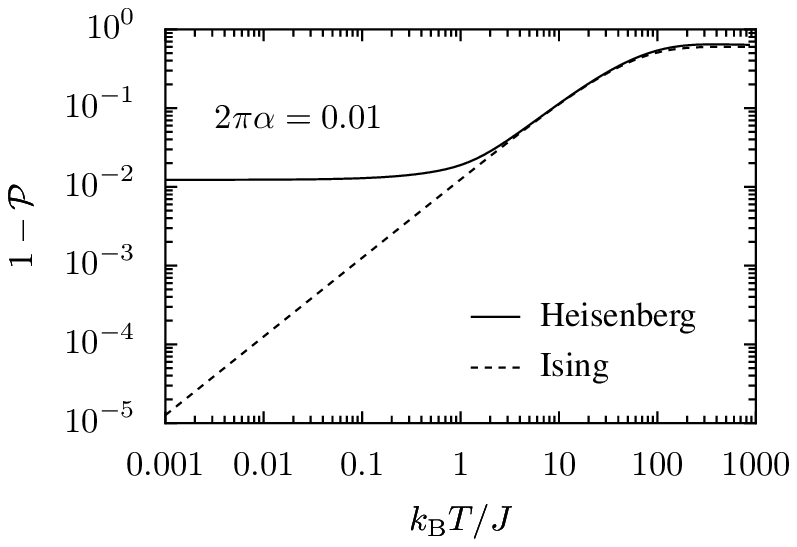}%
\vchcaption{Purity loss $(1-\mathcal{P})$ of two qubits interacting for
a time $t_J = \pi\hbar/4J$ comparing Heisenberg (solid) and Ising
(dashed) coupling.
The dimensionless dissipation strength is $\alpha=0.01/2\pi$.
\label{fig:ising}}
\end{vchfigure}

It is intriguing to compare the result \eqref{GammaH} with the
purity decay of a similar system, namely a pair of qubits
interacting with the Ising interaction
\begin{equation}
\label{Hising}
H_\mathrm{Ising} = J \sigma_1^x \sigma_2^x .
\end{equation}
Such an interaction can also be employed for the construction of
quantum gates like the CNOT gate in Fig.~\ref{fig:cnot}b.  (Note,
however, that for several physical realizations of a quantum computer
\cite{Loss1998a, Kane1998a}, the natural interaction is of the
Heisenberg type).  The main difference to the Heisenberg interaction
\eqref{Hheisenberg}, is the fact that $H_\mathrm{Ising}$ commutes with
the bath coupling \eqref{Hcoupl}, i.e., the bath couples to a good
quantum number.  Carrying out the same calculation as above yields the
purity decay
\begin{equation}
\label{purity-ising}
\Gamma_\mathrm{Ising} = \frac{4}{5}\mathcal{S}(0) =
\frac{8}{5}\pi\alpha k_\mathrm{B}T
\end{equation}
which exhibits a significantly different low-temperature behavior:
Instead of saturating, it remains proportional to the temperature,
cf.\ Fig.~\ref{fig:ising}.

\subsection{Coherence stabilization by an AC field}

In Sections \ref{CDT} and \ref{sec:ising} we presented two results
that lead to the central idea for coherence stabilization of a qubit
pair under the influence of a Heisenberg coupling: First, an AC field
can suppress the dynamics ``transverse'' to the driving, i.e., it can
effectively eliminate the parts of the Hamiltonian that depend on
spin matrices other than $\sigma_1^x$.  Second, qubits with Ising
interaction are less sensitive to decoherence than qubits with
Heisenberg interaction.  Thus, the question arises whether one can
act with an AC field on the system \eqref{H0} in such a way that
precisely the part of the Hamiltonian \eqref{Hheisenberg} that causes
the quantum noise becomes suppressed.  This is indeed the case and can
be performed by driving qubit~1 with the AC field \eqref{HF:CDT}.
Note that qubit 2 remains undriven.

For the computation of the coherence properties, we proceed as in
Section \ref{CDT}: We first derive an effective static qubit
Hamiltonian by transforming the time-dependent Hamiltonian
$H_\mathrm{Heisenberg} + H_\perp(t)$ via \eqref{Uac} into rotating
frame and subsequently replace it by its time average.  After some
algebra along the lines of Section \ref{CDT}, we obtain the
Hamiltonian
\begin{equation}
\label{Heff}
\bar H_\mathrm{qubits} = (J - J_\perp) \sigma_1^x\sigma_2^x
+ J_\perp \, \vec\sigma_1\cdot \vec\sigma_2 ,
\end{equation}
where the constant
\begin{equation}
J_\perp = J \langle \cos[2\phi(t)] \rangle_{2\pi/\Omega}
\end{equation}
denotes an effective interaction ``transverse'' to the driving and
$\langle\ldots\rangle_{2\pi/\Omega}$ the time average over the driving
period.  Consequently, we find the S-matrix $S(t,t') =
\exp\{-\mathrm{i} \bar H_\mathrm{qubits}(t-t')/\hbar\}$, such that the
propagator of the \textit{driven} system again assumes the form
\eqref{U(t,t')} with $\bar H_\mathrm{qubit1}$ replaced by $\bar
H_\mathrm{qubits}$.
Having this propagator at hand, we are in the position to derive explicit
expressions for the operators $\sigma_j^x(t-\tau,t)$ and $Q_j$.
Again, the calculation is conveniently done in the basis of the total
spin $\vec L$ and $L_x$ which, owing to the relation $\sigma_1^x
\sigma_2^x = \frac{1}{2}(\sigma_1^x+\sigma_2^x)^2-1$, is an eigenbasis of
the Hamiltonian \eqref{Heff}.  We evaluate the purity decay
rate~\eqref{puritydecay} in this basis and finally obtain
\begin{equation}
  \label{purity2}
  \Gamma_\mathrm{Heisenberg,driven}
  = \frac{2}{5}\{ \mathcal{S}(0) + \mathcal{S}(4J_\perp/\hbar) \} ,
\end{equation}
i.e., the result \eqref{GammaH} but with $J$ replaced by $J_\perp$.
For $f(t)\equiv 0$, we find $J_\perp=J$ such that the static result is
reproduced; otherwise, the inequality
$|J_\perp| < J$ holds and, thus, the bath correlation function
$\mathcal{S}$ in Eq.~\eqref{purity2} has to be evaluated at a lower
frequency.  For an ohmic or a super-ohmic bath, $\mathcal{S}(\omega)$ is a
monotonously increasing function and, consequently, the AC field reduces
purity decay (unless $J > \omega_\mathrm{cutoff}$).

The purity decay assumes its minimum for $J_\perp=0$.  This condition marks
the working points on which we shall focus henceforth.  For an ohmic
spectral density $I(\omega) = 2\pi\alpha\omega$, the purity decay at the
working points becomes $\Gamma = \frac{4}{5} \mathcal{S}(0) =
8\pi\alpha k_\mathrm{B}T/5$.  This value has to be
compared to the purity decay in the absence of driving: An analysis reveals
that for $k_\mathrm{B}T>J$, decoherence is essentially driving independent.  By
contrast for low temperatures, $k_\mathrm{B}T<J$, the driving reduces the
decoherence rate by a factor $k_\mathrm{B}T/2J$.  This low-temperature behavior
results from the fact that for $J_\perp=0$, the effective Hamiltonian
\eqref{Heff} is identical with the Ising Hamiltonian \eqref{Hising}
and, thus, commutes with the qubit-bath coupling operators
$\sigma_j^x$.  Note that the latter are not affected by the
transformation \eqref{Uac}.  This means that the driving modifies the
effective qubit Hamiltonian such that the bath acts as pure phase noise whose
influence is proportional to the temperature.
The fact that the driving field couples to the same coordinate as the bath
distinguishes the present coherence stabilization from dynamical
decoupling.  In that respect, the present scheme is complementary to
coherence-preserving qubits \cite{Bacon2001a}, for which heating
errors are the only source of decoherence.

For a rectangular driving for which $f(t)$ switches between the values
$\pm A/2$, the condition $J_\perp=0$ is equivalent to $A=\hbar\Omega$
and corresponds to two $\pi$-pulses per period.
For a harmonic driving, $f(t)=A\cos(\Omega t)/2$, one obtains $J_\perp =
J J_0(A/\hbar\Omega)$, where $J_0$ denotes the zeroth-order Bessel function
of the first kind.  Then, at the working points $J_\perp=0$, the ratio
$A/\hbar\Omega$ assumes a zero of $J_0$, i.e., one of the values 2.405..,
5.520.., 8.654..,~\ldots.

So far, we ignored that the driving also affects the coherent dynamics
and, thus, the pulse sequence of the CNOT operation needs a
modification:
At the working points of the driven system, the propagator
becomes $U_\mathrm{eff}(t,t') = U_\mathrm{I}(J(t-t')/\hbar)$
where
\begin{equation}
\label{UI}
U_\mathrm{I}(\varphi) = \exp(-\mathrm{i}\varphi \sigma_1^x\sigma_2^x )
\end{equation}
is the propagator corresponding to the ideal Ising interaction \eqref{Hising}.
This allows one to implement the alternative CNOT operation depicted in
Fig.~\ref{fig:cnot}b \cite{Gershenfeld1997a, Makhlin2001a,
Galindo2002a}.  Note that the interaction time $t_J = \pi\hbar/4 J$ is
the same as for the original gate operation in Fig.~\ref{fig:cnot}a.
Since $U_\mathrm{AC}(2\pi/\Omega)$ is the identity [cf.\
Eq.~\eqref{Uac}], we assume for convenience that $t_J$ is an integer
multiple of the driving period $2\pi/\Omega$, i.e., $\Omega =
8kJ/\hbar$ with integer $k$ .

\subsection{Numerical solution}
\begin{vchfigure}[tb]
\includegraphics{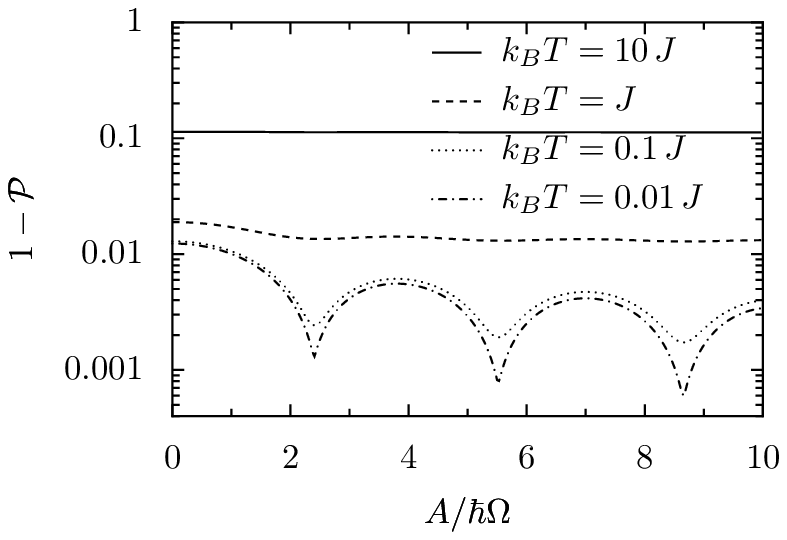}
\vchcaption{Purity loss during the interaction time
$t_J$ as a function of the driving amplitude.
The driving frequency is $\Omega=32J/\hbar$ and the dissipation strength
$2\pi\alpha=0.01$.  For $A=0$, the undriven situation is reproduced.
\label{fig:purity.A}}
\end{vchfigure}%
In order to confirm our analytical results, we compute the
dissipative propagator as described in Section \ref{sec:numMethod}.
Thereby, we restrict ourselves to purely harmonic driving
$f(t)=A\cos(\Omega t)/2$.  The resulting purity loss during the
interaction time $t_J$ is depicted in Fig.~\ref{fig:purity.A}.  We
find that for large temperatures, $k_\mathrm{B}T>J$, decoherence is
fairly independent of the driving.  This behavior changes as the
temperature is lowered: Once $k_\mathrm{B}T<J$, the purity loss is
significantly reduced whenever the ratio $A/\hbar\Omega$ is close to a
zero of the Bessel function $J_0$.  Both observations confirm the
preceding analytical estimates.
The behavior at the first working point $A\approx 2.4\,\hbar\Omega$ is
depicted in Fig.~\ref{fig:purityfidelity.omega}a.  For relatively
low driving frequencies, we find the purity loss being proportional to
$J/\Omega$.  This significant deviation from the analytical result for
small $\Omega$ relates to the fact that the low-frequency regime is
not within the scope of our rotating-wave approximation which assumes
$\Omega$ to be the largest frequency scale.  With
increasing driving frequency, the discrepancy decreases until finally
decoherence is dominated by thermal noise $\propto T$ and the
numerical solution confirms the analytical results.
Figure \ref{fig:purity.kT} reveals that the accuracy of our analytical
estimates increases with the driving frequency: While for the relatively
low frequency $\Omega = 16J/\hbar$, the purity loss is still close to
the one of the undriven Heisenberg gate, it converges in the limit
$\Omega \to\infty$ to the value obtained for Ising interaction (cf.\
the dashed line in Fig.~\ref{fig:ising}).
\begin{vchfigure}[tb]
\includegraphics{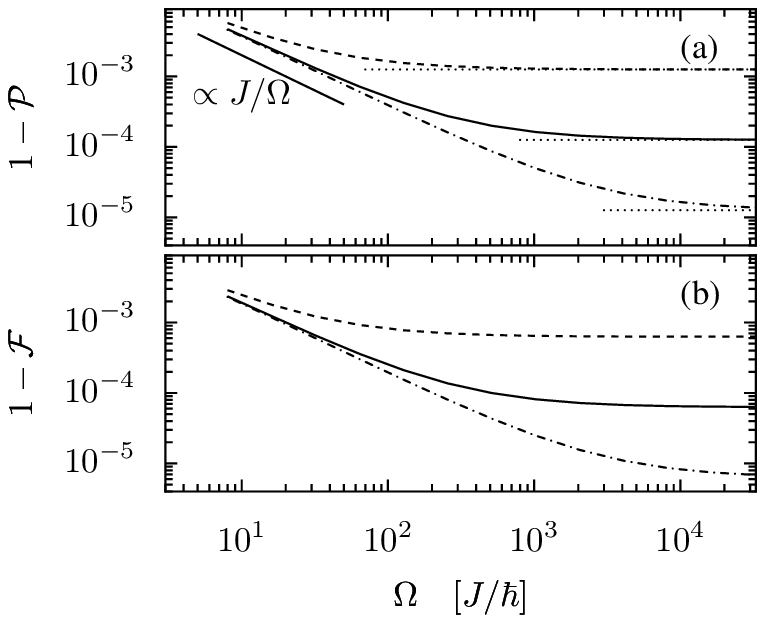}
\vchcaption{(a) Purity loss $(1-\mathcal{P})$ for a pair of qubits with
Heisenberg interaction as a function of the driving frequency for the
temperatures $k_\mathrm{B}T=0.1J$ (dashed), $0.01J$ (solid), and
$0.001J$ (dash-dotted).  The dotted lines mark the analytical estimate
$1-\mathcal{P}(t_J) \approx \Gamma_\mathrm{Ising} t_J$.
(b) Corresponding fidelity defect $1-\mathcal{F}$.
\label{fig:purityfidelity.omega}}
\end{vchfigure}%
\begin{vchfigure}[tb]
\includegraphics{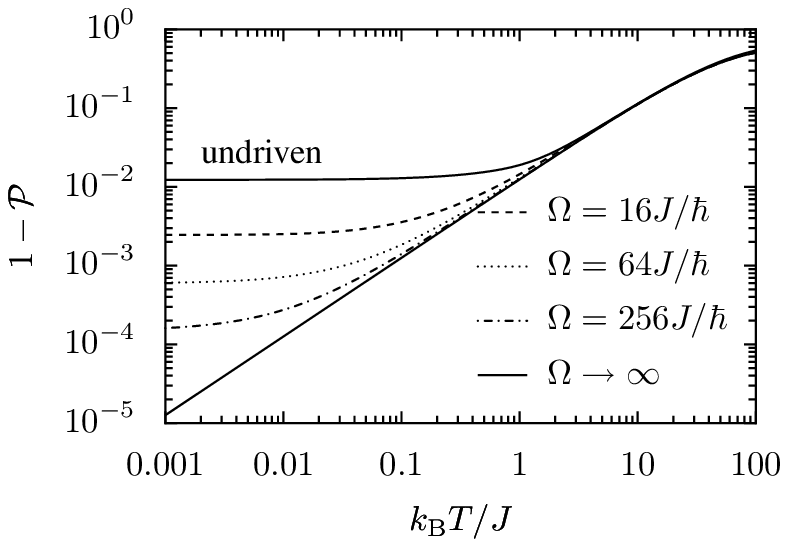}
\vchcaption{Purity loss shown in Fig.~\ref{fig:purityfidelity.omega}
as a function of the temperature.  The driving amplitude $A
\approx 2.4\,\hbar\Omega$ is adjusted such that $1-\mathcal{P}$
assumes its first minimum; cf.\ Fig.~\ref{fig:purity.A}.
All other parameters are as in Fig.~\ref{fig:purityfidelity.omega}.
\label{fig:purity.kT}}
\end{vchfigure}%

Still, there remains one caveat:
The gate operation in Fig.~\ref{fig:cnot}b relies on the fact that
the static effective Hamiltonian $\bar H_\mathrm{qubits}$ describes
the dynamics of the driven system sufficiently well---any
discrepancy results in a coherent error.  Therefore, we still have to
justify that such coherent errors are sufficiently small.
As a measure, we employ the so-called fidelity \cite{Poyatos1997a}
\begin{equation}
\mathcal{F}
= \overline{\operatorname{tr}[\rho_\mathrm{ideal}\,\rho(t_J)]} ,
\end{equation}
which is defined as the overlap between the real outcome of the
operation, $\rho(t_J)$, and the desired final state
$\rho_\mathrm{ideal} = U_\mathrm{I}(\pi/4) \rho_\mathrm{in}
U_\mathrm{I}^\dagger(\pi/4)$ in the average over all pure initial
states.  The time-evolution \eqref{UI} of the ideal Ising qubit-qubit
interaction is characterized by $\mathcal{F}=1$.
Figure \ref{fig:purityfidelity.omega}b demonstrates that
the fidelity defect $1-\mathcal{F}$ at the first working point is even
smaller than the purity loss.  Thus, we can conclude that coherent
errors are not of a hindrance.

\subsection{Implementation with quantum dots}

Figure \ref{fig:purity.kT} indicates that the benefits of a
``preferably infinitely'' large driving
frequency and amplitude.  Thus, a crucial question is whether
sufficiently large values are experimentally within reach.
For spin qubits in quantum dots \cite{Loss1998a} a typical exchange
coupling is $J=0.1\,\mathrm{meV}$ which for a temperature
$T=10\,\mathrm{mK}$ corresponds to the solid lines in
Figs.~\ref{fig:purity.A} and \ref{fig:purityfidelity.omega}.  These
results demonstrate that driving with the feasible frequency $\Omega =
2\pi\times 100\,J/\hbar \approx 10^{12}\,\mathrm{Hz}$  and amplitude
$A=10\,\mathrm{meV}$ already reduces the purity loss by two orders of
magnitude while the fidelity loss stays at a tolerable level.

\section{Conclusions}

We have investigated the influence of oscillating fields on the coherence
properties of one- and two-qubit gate operations for three different
cases for which it is beneficial.
The first case constitutes a continuous-wave version of dynamical
decoupling of a single qubit from its environment.  It is
characterized by a driving Hamiltonian that does not commute with the
bath coupling.  There, we have found that a low-frequency driving is
rather destructive because it generally even enforces decoherence.
However, once the frequency exceeds the bath cutoff, the coherence
properties recover and are finally significantly improved, especially
at high temperatures.  Since such a dynamical decoupling by a harmonic
driving allows higher driving frequencies than the pulsed version,
this form of coherence stabilization bears interesting perspectives
for applications.

A second possibility for manipulating the decoherence of a single qubit is
provided by the physics of coherent destruction of tunneling.  For
such a driving,
we have found that the coherence stabilization results from the fact
that the driving shifts the coherent long-time dynamics of the qubit
towards lower frequencies.  There, the spectral density of an ohmic bath is
lower and, consequently, the effective dissipation is weaker.  This
implies that decoherence is generally reduced---most significantly at
low temperatures.  For high temperatures, however, the lower spectral
density is counterbalanced by an increasing thermal noise, such that
in this regime decoherence is essentially not influenced the driving.

For two qubits interacting via a Heisenberg exchange coupling, a
suited AC field turns the interaction into an effective Ising
interaction, which is much less sensitive to decoherence.
For qubits with such a Heisenberg interaction, like e.g.\ spin qubits, this
suggests the following coherence stabilization protocol: Use for the
CNOT operation a pulse sequence that is suitable for Ising interaction
with the latter being realized by a Heisenberg interaction with a
proper additional AC field.
This coherence stabilization scheme differs from previous proposals in
two respects: First, it is different from dynamical decoupling because
the driving commutes with the bath coupling.  By contrast, the central
idea of our scheme is to suppress rather the {coherent} system
dynamics ``transverse'' to this sensitive system coordinate.
Consequently, the bit-flip noise acts as pure phase noise, which is
proportional to the temperature.  Cooling, thus, enables a further
coherence gain.
The second difference is that the proposed scheme eliminates also the
noise stemming from the spectral range above the driving frequency
and, thus, is particularly valuable for ohmic noise spectra with large
cutoff frequencies.
Moreover, the driven system still allows one to perform the desired
CNOT operation with high fidelity and within the same operation time
as in the absence of the control field.  Hence, the gained coherence
time fully contributes to the number of feasible gate operations.

\begin{acknowledgement}
We acknowledge helpful discussions with K.~M. Fonseca-Romero,
C.~M. Wubs, R. Doll, F.~K. Wilhelm, M.~J. Storcz, and U. Hartmann.
This work has been supported by the Freistaat Bayern via the
network ``Quanteninformation l\"angs der A8''
and by the Deutsche Forschungsgemeinschaft through SFB 631.
\end{acknowledgement}

\appendix
\section{Average over all pure states}
\label{app:average}

In this appendix, we derive formulas for the evaluation of expressions
of the type $\operatorname{tr}(\rho A)$ and $\operatorname{tr}(\rho
A\rho B)$ in an ensemble
average over all pure states $\rho=|\psi\rangle\langle\psi|$.  The state
$|\psi\rangle$ is an element of an $N$-dimensional Hilbert space.
Decomposed into an arbitrary orthonormal basis set
$\{|n\rangle\}_{n=1\ldots N}$, it reads
\begin{equation}
|\psi\rangle = \sum_{n=1}^N c_n |n\rangle ,
\end{equation}
where the only restriction imposed on the coefficients $c_n$ is the
normalization $\langle\psi|\psi\rangle = \sum_n |c_n|^2=1$.  Hence the
ensemble of all pure states is described by the distribution
\begin{equation}
\label{Pc}
P(c_1,\ldots,c_N)
= \frac{2\pi^N}{(N-1)!} \delta\Big(1-\sum_{n=1}^N |c_n|^2\Big) .
\end{equation}
The prefactor on the right-hand side of Eq.~\eqref{Pc} has been
determined from the normalization
\begin{equation}
\int \mathrm{d}^2c_1 \ldots \mathrm{d}^2c_N\, P(c_1,\ldots,c_N) = 1
\end{equation}
of the distribution and $\int \mathrm{d}^2c$ denotes integration over
the real and the imaginary part of $c$.
We emphasize that $P(c_1,\ldots,c_N)$ is invariant under unitary
transformations of the state $|\psi\rangle$.
The computation of the ensemble averages of the coefficients with the
distribution \eqref{Pc} is straightforward and yields
\begin{align}
\overline{c_m c_n^*} = {}& \frac{1}{N}\delta_{mn} , \label{c2}
\\
\overline{c_m c_n^* c_{m'} c_{n'}^*}
= {}& \frac{1}{N(N+1)}(\delta_{mn}\delta_{m'n'} + \delta_{mn'}\delta_{nm'}) .
\label{c4}
\end{align}
Using these expressions, we consequently find for the ensemble averages of
the expressions $\operatorname{tr}(\rho A)$ and $\operatorname{tr}(\rho A\rho B)$ the
results
\begin{align}
\overline{\operatorname{tr}(\rho A)}
={} & \overline{\langle\psi|A|\psi\rangle}
=   \frac{\operatorname{tr}(A)}{N} ,
\\
\overline{\operatorname{tr}(\rho A\rho B)}
={} & \overline{\langle\psi|A|\psi\rangle \langle\psi|B|\psi\rangle}
=   \frac{\operatorname{tr}(A) \operatorname{tr}(B)
          + \operatorname{tr}(AB)}{N(N+1)} ,
\end{align}
which have been used to obtain the purity decay rate \eqref{puritydecay}.


\end{document}